# Reference-State Bootstrap Correction and Kernel Characterization of Residual-Based Fit Indices in Structural Equation Models

Eunjeong Song · Sehee Hong
Department of Education, Korea University, Seoul, Republic of Korea

Eunjeong Song, ORCID iD: https://orcid.org/0000-0002-2302-3227
Sehee Hong, ORCID iD: https://orcid.org/0000-0001-5468-8398

**Abstract**

Sampling error affects the finite-sample values of residual-based fit indices. We show that the unweighted comparative fit index ($\mathrm{CFI}_u$) rests on a quadratic-kernel maximum mean discrepancy and that a Gaussian-kernel comparative fit index (KCFI) converges to $\mathrm{CFI}_u$ as bandwidth increases. Because kernel discrepancies lack an analytic sampling correction, we extend expected-discrepancy bootstrap methods to a fitted-correlation reference state: the sample is transformed to reproduce the fitted correlation matrix, then resampled and refitted to estimate the exact-fit sampling component. Across 18 confirmatory factor analysis conditions, this correction reduced the root mean square error (RMSE) of both indices by 62%–87%. Under normality it matched an analytic residual-covariance estimator of $\mathrm{CFI}_u$ within .0001, and in small samples it outperformed ordinary bootstrap correction in accuracy and availability. Corrected KCFI had lower RMSE than identically corrected $\mathrm{CFI}_u$ in all 29 nonnormal conditions, even against a common target, but in six heavy-tailed small-sample conditions the analytic estimator was more accurate than both. The results validate the reference-state correction where it is needed and delimit where the analytic estimator is preferable.

*Keywords:* comparative fit index; finite-sample correction; kernel method; maximum mean discrepancy; residual correlation

## Introduction

Evaluating a structural equation model (SEM) requires assessing how well its covariance structure reproduces the data. Because the $\chi^2$ test assesses exact fit and is sensitive to sample size, researchers also use indices that summarize the magnitude of misfit (Browne & Cudeck, 1993; Hu & Bentler, 1999). These global summaries do not identify which associations remain unexplained. Individual residual correlations—the differences between sample and model-implied correlations—locate such discrepancies (Browne et al., 2002; Maydeu-Olivares & Shi, 2017). For example, shared item content or a common response method that the model does not represent produces residual correlations and can distort structural estimates (Cole et al., 2007; Hong et al., 2011; Marsh, 1996). Relative fit indices and individual residual correlations therefore provide complementary information (McDonald & Ho, 2002).

Absolute indices such as RMSEA (Steiger, 1990; Steiger & Lind, 1980) and SRMR (Bentler, 1995) summarize analysis-model misfit; relative indices such as CFI (Bentler, 1990) and TLI (Tucker & Lewis, 1973) compare it with a baseline model (Bentler & Bonett, 1980; Widaman & Thompson, 2003). Maximum likelihood (ML) and unweighted residual criteria weight residual directions differently, so their summaries need not agree (Browne et al., 2002). Lubbe (2023) defined an unweighted approximation-error target—the population sum of squared residual correlations—and the corresponding $\mathrm{RMSEA}_u$ and $\mathrm{CFI}_u$, with estimators that use either the estimated covariance matrix of the residual correlations or the noncentral $\chi^2$ distribution of a test statistic. Both conventional and residual-based indices remain global summaries.

The first question concerns the connection between residual and kernel discrepancies. We show that the squared Frobenius correlation misfit equals the squared maximum mean discrepancy ($\mathrm{MMD}^2$) under a homogeneous quadratic kernel, an instance of moment comparison by polynomial kernels (Gretton et al., 2012; Sriperumbudur et al., 2011). Replacing the quadratic kernel by a Gaussian kernel defines the kernel comparative fit index (KCFI). Its population value depends on the data distribution and the bandwidth, and $\mathrm{CFI}_u$ is its large-bandwidth limit. Because the nonnormal data common in psychology (Cain et al., 2017) may affect a Gaussian-kernel and a quadratic-kernel discrepancy differently, we compare the estimation errors of the two indices against their respective population values and, as a sensitivity analysis, against the common $\mathrm{CFI}_u$ target.

The second question concerns sampling-error correction. CFI subtracts the exact-fit sampling component of the discrepancy, conventionally $df/(n-1)$ in discrepancy-function units. For $\mathrm{CFI}_u$, analytic corrections based on the asymptotic covariance matrix of the residual correlations are available (Lubbe, 2023; Maydeu-Olivares, 2017); for a kernel discrepancy, no such correction exists. Bootstrap methods have been used to estimate exact-fit reference distributions and expected discrepancies (Bollen & Stine, 1992; Kim & Millsap, 2014; Yuan & Hayashi, 2003; Zhang & Savalei, 2016). In particular, Yuan and Marshall (2004) defined misfit as the difference between the expected test statistics under the analysis and null models and estimated it by bootstrap. We extend this expected-discrepancy approach to paired kernel discrepancies and a comparative ratio. A Bollen–Stine-type linear transformation first makes the sample correlation matrix equal the fitted matrix (Beran & Srivastava, 1985; Yuan et al., 2007; Yung & Bentler, 1994); we call the transformed sample the *fitted-correlation reference state*. Resampling and refitting this reference state estimates the sampling component to subtract from the analysis- and baseline-model discrepancies. Because the same procedure applies to the unweighted discrepancy, it can be validated against the analytic correction of $\mathrm{CFI}_u$ and compared with a diagonal-deleted (paired U-form) estimator and with ordinary bootstrap bias correction.

The research questions are:

1. How is the unweighted residual discrepancy represented by a quadratic-kernel MMD², and under what conditions does the Gaussian-kernel KCFI reduce to $\mathrm{CFI}_u$?
2. How much does reference-state correction reduce the finite-sample error of KCFI and $\mathrm{CFI}_u$, relative to analytic, paired U-form, and ordinary bootstrap alternatives, under normal, moderately nonnormal, and heavy-tailed data?

## Kernel Characterization of Residual-Based Fit

### *Definitions and the Unweighted Target*

Throughout, a discrepancy is a nonnegative quantity that measures the difference between two correlation structures or two distributions; the ML discrepancy function is one such quantity. A kernel $k(x, y)$ is a symmetric positive-semidefinite similarity function on pairs of observations. The squared maximum mean discrepancy between two distributions is the squared distance between their mean embeddings in the feature space of $k$ (Gretton et al., 2012; Muandet et al., 2017): for mutually independent copies $X, X' \sim P$ and $Y, Y' \sim Q$,

$$\mathrm{MMD}^2(P, Q; k) = \mathbb{E}\{k(X, X')\} + \mathbb{E}\{k(Y, Y')\} - 2\mathbb{E}\{k(X, Y)\}. \qquad (1)$$

A kernel is characteristic when this quantity is zero only if $P = Q$ (Sriperumbudur et al., 2010). The bandwidth $h$ of the Gaussian kernel $k_h(x, y) = \exp\{-\|x - y\|^2/(2h^2)\}$ is its length scale.

Let $n$ be sample size, $p$ the number of measured variables, $Z \in \mathbb{R}^{n \times p}$ the standardized data, and $R = n^{-1} Z^\top Z$ the sample correlation matrix. We use denominator-$n$ covariances and $T = nF_{ML}$, the convention also used by lavaan (Rosseel, 2012); the corresponding normal-theory first-order sampling correction is $df/n$. Common scaling factors cancel within a comparative ratio, but the covariance and test-statistic conventions must be used consistently. The standardized population distribution $P$ has correlation matrix $R_P$. The model-implied correlation matrix is $R_M^*$ in the population and $\hat{R}_M$ in a sample, or $R_M$ when no distinction is needed. The model is fitted by ML to the sample covariance matrix $S$, with discrepancy function

$$F_{ML}(S, \Sigma) = \log|\Sigma| - \log|S| + \mathrm{tr}(S\Sigma^{-1}) - p. \qquad (2)$$

Under misspecification, $R_M^*$ corresponds to the ML pseudo-true parameter, the probability limit of the ML estimator (White, 1982). With population ML discrepancy values $F_M^*$ and $F_B^*$ for the analysis and baseline models, $\mathrm{CFI}^* = 1 - \max(F_M^*, 0)/\max(F_M^*, F_B^*, 0)$.

Let $E = R_P - R_M$ be the residual correlation matrix, with $E_M$ and $E_B = R_P - I_p$ for the analysis and baseline models. Lubbe's (2023) population unweighted comparative fit index is

$$\mathrm{CFI}_u^* = 1 - \frac{\|E_M\|_F^2}{\max\{\|E_M\|_F^2, \|E_B\|_F^2\}} = \max\left\{1 - \left(\frac{\mathrm{SRMR}_M}{\mathrm{SRMR}_B}\right)^2, 0\right\}, \qquad (3)$$

for a positive baseline discrepancy, with the same SRMR normalization for both models. We call the zero-correlation comparison model the *baseline model* because zero correlation does not imply independence under nonnormal distributions.

### *Exact Quadratic-Kernel Representation*

*Proposition 1 (exact quadratic-kernel representation).* If $P$ and $Q$ are centered with covariance matrices $R_P$ and $R_Q$, then for the homogeneous quadratic kernel $k_2(x, y) = (x^\top y)^2$,

$$\mathrm{MMD}^2(P, Q; k_2) = \mathrm{tr}(R_P^2) + \mathrm{tr}(R_Q^2) - 2\mathrm{tr}(R_P R_Q) = \left\|R_P - R_Q\right\|_F^2. \qquad (4)$$

The proof (Section S1.2) uses only finite second moments. Hence, if $Q_M$ and $Q_B$ are any centered distributions with covariance matrices $R_M$ and $I_p$, normalizing their quadratic-kernel discrepancies from $P$ gives $\mathrm{CFI}_u^*$ exactly. Because $k_2$ is not characteristic on general distributions (Sriperumbudur et al., 2011), zero discrepancy means equal second moments, not equal distributions.

***Fitted-Correlation Transformation and Population KCFI***

A linear transformation carrying $R$ to $\hat{R}_M$ is not unique. We use

$$A_M = R^{-1/2}\left(R^{1/2}\hat{R}_M R^{1/2}\right)^{1/2} R^{-1/2}, \qquad (5)$$

the unique symmetric positive-definite matrix with $ARA^\top = \hat{R}_M$; among all such matrices it minimizes $\mathbb{E}\|AX - X\|^2$, the mean squared movement of the observations, and it is the optimal transport map between normal distributions with these covariance matrices (Dowson & Landau, 1982; Olkin & Pukelsheim, 1982). The transformed data $Z_M = ZA_M$ satisfy $n^{-1}Z_M^\top Z_M = \hat{R}_M$: their empirical distribution reproduces the fitted correlation structure but not the full distribution implied by the SEM. For the baseline model, $A_B = R^{-1/2}$ and $Z_B = ZA_B$ has correlation matrix $I_p$. Because the Gaussian kernel depends on Euclidean distances, the data are standardized and the analysis is restricted to scale-invariant covariance-structure models (Cudeck, 1989); $R$ must be positive definite, so $n > p$ is required.

Let $A_M^*$ and $A_B^*$ be the population transformations and $Q_M^*$ and $Q_B^*$ the distributions of $A_M^*X$ and $A_B^*X$ for $X \sim P$. With population bandwidth $h_P$, define $D_M^* = \mathrm{MMD}^2(P, Q_M^*; k_{h_P})$ and $D_B^* = \mathrm{MMD}^2(P, Q_B^*; k_{h_P})$. When $D_B^* > 0$, the population KCFI is

$$\mathrm{KCFI}^* = 1 - \frac{D_M^*}{\max(D_M^*, D_B^*)}. \qquad (6)$$

Under correct specification $A_M^* = I_p$ and $\mathrm{KCFI}^* = 1$; if $R_P = I_p$, the index is undefined. Because $Q_M^*$ is a linear image of $P$, it differs from $P$ only through its correlation structure. KCFI therefore does not measure distributional misfit; it measures, on the kernel scale, how far the data distribution must be moved for its correlation structure to agree with the model. Its population value depends on $P$ beyond $R_P$, but this dependence was small in the populations studied here (Table S2). Figure 1 illustrates the construction. In a sample,

$$\hat{D}_M = \frac{1}{n^2}\sum_{i,j} k_h\,(z_i, z_j) + \frac{1}{n^2}\sum_{i,j} k_h\,(z_{Mi}, z_{Mj}) - \frac{2}{n^2}\sum_{i,j} k_h\,(z_i, z_{Mj}), \qquad (7)$$

and $\hat{D}_B$ replaces $Z_M$ by $Z_B$. This empirical discrepancy has V-statistic form, including same-observation terms ($i = j$); the *paired U-form (diagonal-deleted) estimator* introduced below excludes them. The bandwidth $h$ is the median of the positive pairwise Euclidean distances in $Z$, the median heuristic (Garreau et al., 2018; Gretton et al., 2012), and is held fixed while a sample is evaluated.

**Figure 1**

*Construction of KCFI From the Fitted-Correlation Linear Transformation and MMD² Discrepancies*

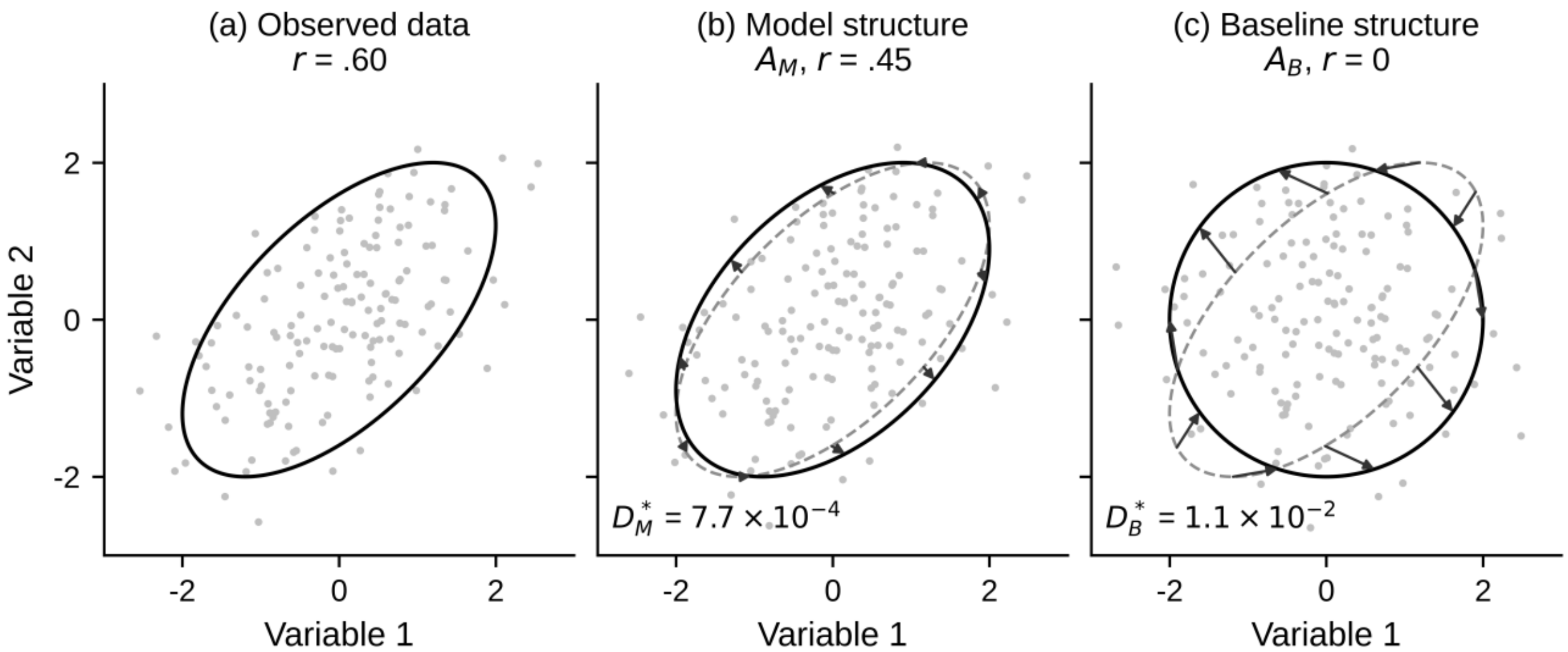

*Note.* Panel (a) shows 150 standardized observations drawn from a two-variable normal population with correlation .60 and the population correlation ellipse. Panels (b) and (c) show the minimum-movement transformations $A_M$ and $A_B$ carrying the data to the analysis-model (correlation .45) and baseline (correlation 0) structures. Dashed and solid ellipses mark the observed and target structures, and arrows show the displacement of corresponding points on the ellipse. $D_M^*$ and $D_B^*$ are the closed-form normal-theory squared maximum mean discrepancies (MMD²) between the population and its transformed image at the median-heuristic bandwidth of the plotted sample ($h = 1.60$), giving $\text{KCFI} = 1 - D_M^*/\max(D_M^*, D_B^*) = .928$ (KCFI = kernel comparative fit index). The indices in the study are computed on $p$-dimensional standardized data.

### ***Large-Bandwidth Limit***

*Theorem 1 (large-bandwidth expansion).* Let $P$ and $Q$ have mean zero, finite fourth moments, and covariance matrices $R_P$ and $R_Q$, and let $E = R_P - R_Q$. Then

$$\text{MMD}^2(P, Q; k_h) = \frac{\{\text{tr}(E)\}^2}{4h^4} + \frac{\|E\|_F^2}{2h^4} + o(h^{-4}). \qquad (8)$$

If $R_P$ and $R_Q$ are correlation matrices, $\text{tr}(E) = 0$ and, by Proposition 1, $\text{MMD}^2(P, Q; k_h) = (2h^4)^{-1}\text{MMD}^2(P, Q; k_2) + o(h^{-4})$. Consequently, when $R_P \neq I_p$, $\text{KCFI}^* \to \text{CFI}_u^*$ as $h \to \infty$, at rate $O(h^{-2})$ if sixth moments are finite (Section S1.3).

Related expansions occur in kernel two-sample testing (Ramdas et al., 2015); here zero means and equal traces leave a leading term determined by the residual correlations alone. Every linear transformation producing the target correlation matrix has the same large-bandwidth limit; the minimum-movement choice affects only finite-bandwidth values. The median-heuristic bandwidth was about 5.21 in the studied $p = 15$ populations, and increasing $n$ under the median heuristic does not drive $h$ to infinity; the finite-bandwidth differences are reported in Table S2.

**Fitted-Correlation Reference-State Correction**

Even under correct specification, sampling and model fitting generally make $\widehat{D}_M$ positive. The empirical distribution of $Z_M$, whose correlation matrix equals $\hat{R}_M$, is the fitted-correlation reference state. Resample its rows with replacement (Efron & Tibshirani, 1993), restandardize, refit the model, and recompute the transformation and discrepancy. The mean over successful discrepancies among $L$ bootstrap replications, $\hat{c}_M$, estimates the sampling component generated by this complete procedure under the reference state. For the baseline, resample $Z_B$, restandardize, and decorrelate again to obtain $\hat{c}_B$. The bandwidth is fixed at its original-sample value. Subtracting this exact-fit reference mean uses the working approximation $\mathbb{E}(\widehat{D}_M) \approx D_M^* + c_M$. Under misspecification the sampling component of the observed state can differ from that of the reference state, and estimating the model-implied correlation adds variability of its own; the fixed-correlation diagnostic of the focused study examines both effects.

The *reference-state correction* yields the corrected index, defined when $\widehat{D}_B > \hat{c}_B$,

$$\mathrm{KCFI} = 1 - \frac{\max(\widehat{D}_M - \hat{c}_M, 0)}{\max(\widehat{D}_M - \hat{c}_M,\ \widehat{D}_B - \hat{c}_B)}. \qquad (9)$$

The index is 1 when $\widehat{D}_M \leq \hat{c}_M$ and 0 when the corrected analysis-model discrepancy is at least the corrected baseline discrepancy. The uncorrected index is $\mathrm{KCFI}^{\mathrm{raw}} = 1 - \widehat{D}_M/\max(\widehat{D}_M, \widehat{D}_B)$. Applying the same resampling procedure to $\widehat{U}_M = \left\|R - \hat{R}_M\right\|_F^2$ and $\widehat{U}_B = \left\|R - I_p\right\|_F^2$ gives correction terms $\hat{c}_M^{(\infty)}$, $\hat{c}_B^{(\infty)}$, and the index $\mathrm{CFI}_u^{\mathrm{ref}}$. In unweighted least-squares units, $F_{ULS} = \widehat{U}/2$, the reference-state correction term is $\hat{c}_M^{(\infty)}/2$, which estimates the same first-order expected squared residuals as the analytic residual-covariance correction (Section S3).

We compare two alternatives. The paired U-form (diagonal-deleted) estimator deletes same-observation kernel terms and can be negative (Section S5.1). It is unbiased for fixed kernels and independently sampled pairs, but not necessarily when standardization, transformation, and bandwidth are estimated from the same data. Fixed-bandwidth *ordinary bootstrap correction* resamples the observed $Z$, refits, and subtracts the estimated bias $\text{mean}_*(\hat{D}^*) - \hat{D}$, yielding $2\hat{D} - \text{mean}_*(\hat{D}^*)$ (Efron, 1979). Thus, ordinary bootstrap starts from the observed empirical distribution, whereas reference-state correction starts from a transformed distribution with zero fitted correlation misfit.

## Methods

Population calculations established the relation between the two indices; primary and focused simulations evaluated finite-sample corrections; an empirical example illustrated reporting. The base population was a three-factor confirmatory factor analysis (CFA) model with 15 measured variables, five per factor, primary loadings .60, and factor correlations .30. Misspecifications were calibrated by numerical root finding to prespecified population CFI values. Data, code, settings, and reference outputs are provided with the study (Data Availability Statement); computing details are in Section S6.

### *Population Analyses*

The two principal misspecified populations contained an omitted cross-loading or an omitted error covariance, each calibrated to population CFI = .980 (cross-loading .262, error correlation .349). A shared-population example compared residual structures at equal CFI and degrees of freedom (Table S1). Gaussian-kernel values were calculated in closed form under normality at 0.2, 0.4, 1, 2, and 4 times the population median bandwidth $h_P$. Distributional sensitivity held $R_P$ and $R_M^*$ fixed across normal, Vale–Maurelli, multivariate $t_5$, and 5% contaminated-normal distributions at common bandwidths (Table S2).

### *Primary Simulation*

The primary simulation crossed sample size ($n = 200{,}500{,}1{,}000$), distribution (multivariate normal; Vale–Maurelli with marginal skewness 2 and excess kurtosis 7; Curran et al., 1996; Vale & Maurelli, 1983), and model specification (correct specification, omitted cross-loading, omitted error covariance), giving 18 conditions with 500 data sets each. Models were estimated by ML, and Satorra–Bentler scaled statistics were computed for the robust estimators (Brosseau-Liard & Savalei, 2014; Satorra & Bentler, 1994).

Each data set yielded the four discrepancies and $L = 100$ bootstrap replications per reference state; a check in the most demanding cells showed that $L = 100$ and $L = 400$ differed by 5.6%–7.0% of the between-data-set standard deviation (Section S4). Corrected values required at least 90% successful replications for both the analysis and baseline procedures; failed convergence, negative variances, boundary solutions, and non-positive-definite matrices were failures. Raw and corrected KCFI were evaluated against distribution-specific population KCFI at the median-heuristic bandwidth. Five estimators were evaluated against $\mathrm{CFI}_u^*$: the raw index, $\mathrm{CFI}_u^{\mathrm{ref}}$, the approximation discrepancy ratio (ADR) estimator $\mathrm{CFI}_u^{\mathrm{ADR}}$ (Lubbe, 2023, approach 2), its Satorra–Bentler-scaled variant $\mathrm{CFI}_u^{\mathrm{rADR}}$, and an adapted residual-covariance estimator $\mathrm{CFI}_u^{\mathrm{asy}}$ based on Lubbe's approach 1.2 (Browne, 1984; Maydeu-Olivares, 2017). Lubbe's approach 1.1, whose unweighted residual projection is matched to unweighted least-squares estimation, was not evaluated because all models here are estimated by ML, for which approach 1.2 is the natural counterpart. Section S3 gives the implemented formulas. Bias, root mean square error (RMSE), availability, and the proportion equal to 1 were summarized with Monte Carlo standard errors (MCSEs; Morris et al., 2019). A 12-condition severity analysis used population CFI = .955 (Section S4).

### *Focused Comparison at Small Sample Sizes*

The focused study comprised 26 conditions with 500 data sets each: 12 core conditions crossing $n = 100{,}200$, normal and Vale–Maurelli distributions, and the three specifications at population CFI = .980; six stress conditions at $n = 100$ with multivariate $t_5$ or 5% contaminated-normal data; four stronger-misfit conditions at $n = 100$ and population CFI = .955; and four diagnostic conditions that repeated the $n = 100$ misspecified normal and Vale–Maurelli cells on the same generated data with the analysis-model correlation fixed at its population pseudo-true value, to isolate the effect of estimating the model-implied correlation.

The four KCFI procedures were raw V-form estimation, the paired U-form estimator, ordinary bootstrap correction, and reference-state correction. Both bootstraps used $L = 100$, the original bandwidth, and common row indices. Raw, ordinary-bootstrap, and reference-state $\mathrm{CFI}_u$ values were evaluated in parallel. Analytic comparators were added by seed-based regeneration;

22 of 10,926 analyzed replications failed the archived numerical agreement check and were excluded from those comparisons (Section S3.2). Pairwise comparisons required both procedures to be defined. Alternative-minus-reference squared-error contrasts used 95% Monte Carlo intervals of $\pm 1.96$ MCSE, and matched estimated- and fixed-correlation cells provided an interaction diagnostic. Additional checks examined the effect of the archived exclusions (Section S3.2), the exact-fit sampling component and its two estimates (Section S5.3), and scoring both corrected indices against the common $\mathrm{CFI}_u^*$ target (Section S5.4).

### *Empirical Illustration*

Three CFA models were fitted to nine ability tests from the Holzinger and Swineford (1939) data distributed with lavaan (Rosseel, 2012): a one-factor model, a correlated two-factor model, and the visual–textual–speed three-factor model. The sample comprised 301 seventh- and eighth-grade students from two American schools. We used $L = 500$ and compared fit indices and residuals descriptively. Section S6 provides further estimates and computing details.

## Results

### *Population Analyses*

Models with equal degrees of freedom and population CFI = .980 had different residual-based fit when residual structures differed ($\mathrm{CFI}_u^* = .9861$ versus .9938; Table S1). Under normality, KCFI approached $\mathrm{CFI}_u^*$ with increasing bandwidth: for omitted cross-loading, the difference changed from $-.0077$ at $0.2h_P$ to $-.0001$ at $4h_P$. From $h_P$ onward, doubling the bandwidth reduced the gap by approximately a factor of four, consistent with the $O(h^{-2})$ rate. At the common normal-distribution bandwidth $h_P$, the across-distribution range was .00100 for omitted cross-loading and .00054 for omitted error covariance; at $2h_P$ these ranges were comparable to the combined MCSE (Table S2).

### *Primary Simulation*

All indices were defined in all 9,000 data sets. Reference-state correction reduced KCFI RMSE in every condition by 63.9%–87.0% (mean 77.7%) and $\mathrm{CFI}_u$ RMSE by 62.1%–86.4% (mean 76.1%; Table 1, Figure 2). Corrected bias ranged from $-.0071$ to $.0000$ for KCFI and from $-.0095$ to $.0000$ for $\mathrm{CFI}_u^{\mathrm{ref}}$; corrected-RMSE MCSEs were .0001–.0008. Section S4 summarizes uncorrected error and boundary behavior, with complete condition-level results in the analysis package.

**Table 1**

*Finite-Sample RMSE of Corrected Estimators Across the 18 Primary Conditions*

| **Distribution** | **Model** | $n$ | **KCFI** | $\mathrm{CFI}_u^{\mathrm{ref}}$ | $\mathrm{CFI}_u^{\mathrm{ADR}}$ | $\mathrm{CFI}_u^{\mathrm{rADR}}$ | $\mathrm{CFI}_u^{\mathrm{asy}}$ |
|---|---|---|---|---|---|---|---|
| Normal | Correct | 200 | .01107 | .01073 | .01025 | .01059 | .01066 |
| Normal | Correct | 500 | .00375 | .00352 | .00310 | .00316 | .00353 |
| Normal | Correct | 1,000 | .00182 | .00172 | .00140 | .00141 | .00169 |
| Normal | Cross-loading | 200 | .01451 | .01405 | .01159 | .01172 | .01404 |
| Normal | Cross-loading | 500 | .00703 | .00666 | .00659 | .00657 | .00666 |
| Normal | Cross-loading | 1,000 | .00476 | .00455 | .00476 | .00475 | .00454 |
| Normal | Error covariance | 200 | .01273 | .01221 | .01268 | .01299 | .01215 |
| Normal | Error covariance | 500 | .00576 | .00547 | .00513 | .00518 | .00546 |
| Normal | Error covariance | 1,000 | .00363 | .00347 | .00318 | .00319 | .00345 |
| Vale–Maurelli | Correct | 200 | .01235 | .01471 | .02084 | .01236 | .01223 |
| Vale–Maurelli | Correct | 500 | .00469 | .00518 | .00827 | .00422 | .00450 |
| Vale–Maurelli | Correct | 1,000 | .00199 | .00210 | .00380 | .00177 | .00195 |
| Vale–Maurelli | Cross-loading | 200 | .01481 | .01641 | .01811 | .01238 | .01480 |
| Vale–Maurelli | Cross-loading | 500 | .00759 | .00791 | .00737 | .00707 | .00765 |
| Vale–Maurelli | Cross-loading | 1,000 | .00514 | .00530 | .00466 | .00564 | .00531 |
| Vale–Maurelli | Error covariance | 200 | .01458 | .01662 | .02249 | .01506 | .01453 |
| Vale–Maurelli | Error covariance | 500 | .00709 | .00748 | .01012 | .00680 | .00700 |
| Vale–Maurelli | Error covariance | 1,000 | .00399 | .00401 | .00500 | .00369 | .00400 |

*Note.* RMSE = root mean square error relative to each index's population target over 500 replications. KCFI = kernel comparative fit index with reference-state correction, evaluated against population KCFI at each distribution's median-heuristic bandwidth; all other columns target $\mathrm{CFI}_u^*$, the population unweighted comparative fit index. $\mathrm{CFI}_u^{\mathrm{ref}}$ = reference-state correction; $\mathrm{CFI}_u^{\mathrm{ADR}}$ = Lubbe's (2023) approximation-discrepancy-ratio estimator; $\mathrm{CFI}_u^{\mathrm{rADR}}$ = its variant using Satorra–Bentler scaled statistics; $\mathrm{CFI}_u^{\mathrm{asy}}$ = adapted residual-covariance estimator (Section S3).

**Figure 2**

*Finite-Sample RMSE of KCFI and* $\mathrm{CFI}_u$ *Before and After Reference-State Correction*

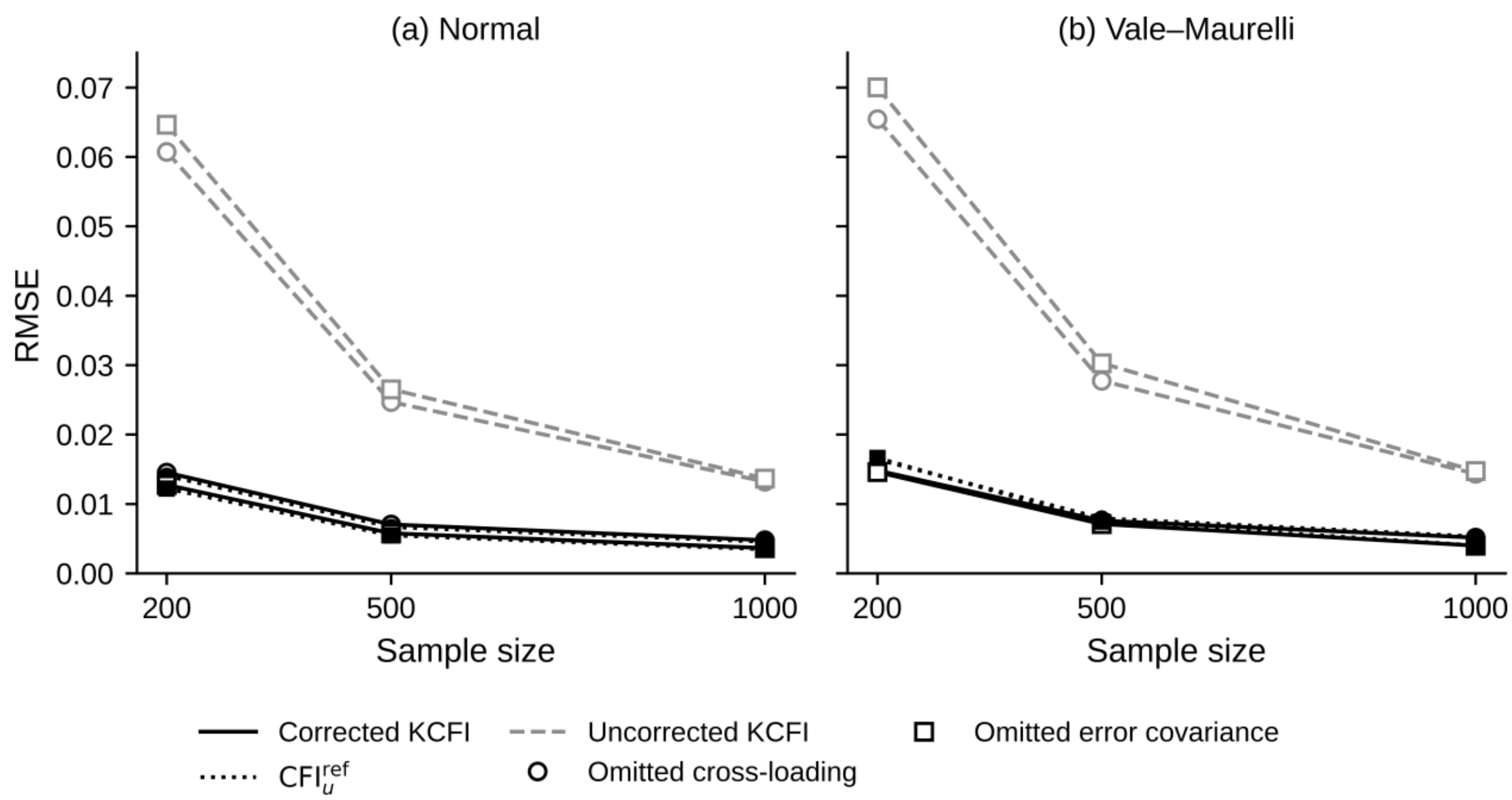

*Note.* RMSE = root mean square error relative to each condition's population value in the misspecified conditions at population CFI = .980, under (a) the multivariate normal and (b) the Vale–Maurelli distribution. Solid lines show corrected KCFI (kernel comparative fit index), dotted lines show $\mathrm{CFI}_u^{\mathrm{ref}}$ (unweighted comparative fit index with reference-state correction), and dashed gray lines show uncorrected KCFI on the same replications. Circles indicate omitted cross-loading and squares omitted error covariance.

In the nine normal conditions, the RMSEs of $\mathrm{CFI}_u^{\mathrm{asy}}$ and $\mathrm{CFI}_u^{\mathrm{ref}}$ differed by less than .0001. Under Vale–Maurelli data, mean RMSE was .0112 for ADR, .0077 for scaled ADR, .0080 for the adapted residual-covariance estimator, and .0089 for reference-state correction. The latter two analytic comparators each outperformed $\mathrm{CFI}_u^{\mathrm{ref}}$ in eight of nine conditions. KCFI had slightly higher RMSE than $\mathrm{CFI}_u^{\mathrm{ref}}$ under normality, but lower RMSE in all nine Vale–Maurelli conditions (0.6%–16.0%). Corrected KCFI equaled 1 in 40.8%–51.2% of correctly specified replications. At population CFI = .955, correction again reduced KCFI RMSE in all 12 conditions (46.6%–71.7%) and $\mathrm{CFI}_u$ RMSE by 43.0%–68.4%, with availability of at least 95.4% (Table S3).

***Focused Comparison***

Analysis-model estimation succeeded in 96.8%–100% of estimated-model replications within each condition. Table 2 gives replication-weighted pooled comparisons, and Tables S5–S6 give condition-level results.

**Table 2**

*Pooled Paired Comparisons in the Focused Study*

*Panel A. Reference-State Correction Versus Alternative Finite-Sample Procedures (22 Estimated-Model Conditions)*

| Index | Alternative | Paired replications | Reference RMSE | Alternative RMSE | RMSE reduction (%) | Conditions with interval above 0 |
|---|---|---|---|---|---|---|
| KCFI | Raw V-statistic | 10,397 | .04546 | .13497 | 66.3 | 22 |
| KCFI | Paired U form | 10,397 | .04546 | .12348 | 63.2 | 22 |
| KCFI | Ordinary bootstrap | 9,386 | .03422 | .04120 | 16.9 | 20 |
| $\mathrm{CFI}_u$ | Raw | 10,397 | .05327 | .12628 | 57.8 | 22 |
| $\mathrm{CFI}_u$ | Ordinary bootstrap | 9,386 | .04041 | .04898 | 17.5 | 22 |

*Panel B. Corrected KCFI Versus Estimators of* $\mathrm{CFI}_u$ *by Condition Group*

| Group | Conditions | Paired replications | KCFI | $\mathrm{CFI}_u^{\mathrm{ref}}$ | $\mathrm{CFI}_u^{\mathrm{ADR}}$ | $\mathrm{CFI}_u^{\mathrm{rADR}}$ | $\mathrm{CFI}_u^{\mathrm{asy}}$ | Conditions with KCFI below $\mathrm{CFI}_u^{\mathrm{asy}}$ |
|---|---|---|---|---|---|---|---|---|
| All estimated | 22 | 10,384 | .04547 | .05328 | .06772 | .04506 | .03276 | 9 |
| Core | 12 | 5,962 | .01800 | .02028 | .02562 | .02064 | .01841 | 6 |
| Stress | 6 | 2,776 | .08090 | .09587 | .12322 | .07899 | .05279 | 0 |
| Stronger misfit | 4 | 1,646 | .02881 | .03037 | .03074 | .02721 | .02904 | 3 |

*Note.* KCFI = kernel comparative fit index; $\mathrm{CFI}_u$ = unweighted comparative fit index; RMSE = root mean square error; Reference = reference-state correction; Paired U form = paired U-form (diagonal-deleted) estimator; Ordinary bootstrap = fixed-bandwidth ordinary bootstrap correction; $\mathrm{CFI}_u^{\mathrm{ref}}$, $\mathrm{CFI}_u^{\mathrm{ADR}}$, $\mathrm{CFI}_u^{\mathrm{rADR}}$, and $\mathrm{CFI}_u^{\mathrm{asy}}$ = reference-state, approximation-discrepancy-ratio, scaled approximation-discrepancy-ratio, and adapted residual-covariance estimators. Each comparison uses only replications in which the compared procedures were defined (in Panel B,

those in which reference-state correction was defined and the archived numerical agreement check was satisfied; Section S3.2). RMSE reduction = $100(\text{RMSE}_{\text{alternative}} - \text{RMSE}_{\text{reference}})/\text{RMSE}_{\text{alternative}}$; Interval above 0 = the 95% Monte Carlo interval of the alternative-minus-reference squared error lies wholly above zero. KCFI is evaluated against population KCFI and the $\text{CFI}_u$ estimators against $\text{CFI}_u^*$, which differ by at most .0019. Core = $n = 100{,}200$ with normal and Vale–Maurelli data; Stress = $n = 100$ with multivariate $t_5$ and contaminated-normal data; Stronger misfit = $n = 100$ at population CFI = .955. Reference-state and ordinary bootstrap corrections were defined in 95.2% and 86.3% of analyzed data sets overall and in 93.6% and 69.2% in the stress conditions.

Reference-state correction had lower KCFI MSE than the paired U-form estimator and ordinary bootstrap correction in all 22 estimated-model conditions. Every condition-level interval favored reference-state correction over the paired U form, and 20 of 22 favored it over ordinary bootstrap; in the other two normal $n = 200$ cells, RMSE differed by less than 1%. The same general pattern occurred for $\text{CFI}_u$, indicating that the improvement was not specific to the kernel. In the stress conditions, reference-state correction reduced paired RMSE relative to ordinary bootstrap by 18.9% for KCFI and 17.8% for $\text{CFI}_u$, and was defined more often (93.6% versus 69.2%). These RMSE comparisons are conditional on both corrections being defined; availability is a separate criterion.

The fixed-correlation diagnostic showed a positive estimated-minus-fixed interaction under normality: .000299, 95% Monte Carlo interval [.000187, .000412], for omitted cross-loading, and .000124 [.000052, .000196] for omitted error covariance (Table S7). In these cells, reference-state correction outperformed ordinary bootstrap only when model correlations were estimated. Under Vale–Maurelli data, an advantage remained with the population model correlation fixed. Thus, model estimation accounted for part, but not all, of the observed difference between corrections.

KCFI had lower RMSE than $\text{CFI}_u^{\text{ref}}$ in all 14 nonnormal focused conditions (10.9%–21.5%) on the ordinary-bootstrap comparison subsets. Across the three studies, this gives 29 nonnormal study conditions, representing 26 distinct design settings because three settings recur. Scoring both indices against $\text{CFI}_u^*$ preserved lower observed KCFI RMSE in all 29; unadjusted 95% Monte Carlo intervals for the paired MSE advantage were wholly positive in 28 (Table S9).

The analytic comparisons limit this advantage (Table 2, Panel B; Table S6). In the 16 core and stronger-misfit conditions, KCFI and $\text{CFI}_u^{\text{asy}}$ were within 7% in RMSE, and KCFI was more accurate in nine. In all six stress conditions, however, $\text{CFI}_u^{\text{asy}}$ had lower RMSE than both reference-corrected indices. Its RMSE was 13%–28% below KCFI in the $t_5$ cells and 44%–46% below it in the contaminated-normal cells. The scaled ADR estimator was better still in two contaminated-normal cells, but unstable in the $t_5$ cells (RMSE up to .120).

### *Empirical Illustration*

**Table 3**

*Global and Residual-Based Fit of Three CFA Models for the Holzinger–Swineford Data*

| Model | $\chi^2$ | $df$ | CFI | TLI | RMSEA [90% CI] | SRMR | Max $\lvert r_{\text{res}} \rvert$ | KCFI | $\text{CFI}_u^{\text{ref}}$ | $\text{CFI}_u^{\text{asy}}$ |
|---|---|---|---|---|---|---|---|---|---|---|
| One-factor | 312.26 | 27 | .677 | .569 | .187 [.169, .206] | .143 | .450 | .730 | .755 | .755 |
| Two-factor (visual–speed / textual) | 151.47 | 26 | .858 | .803 | .127 [.108, .147] | .088 | .285 | .899 | .918 | .917 |
| Three-factor (visual / textual / speed) | 85.31 | 24 | .931 | .896 | .092 [.071, .114] | .065 | .189 | .951 | .959 | .958 |

*Note.* CFA = confirmatory factor analysis; $df$ = degrees of freedom; CFI = comparative fit index; TLI = Tucker–Lewis index; RMSEA = root mean square error of approximation; CI = confidence interval; SRMR = standardized root mean square residual; Max $\lvert r_{\text{res}} \rvert$ = largest absolute residual correlation; KCFI = kernel comparative fit index with reference-state correction ($L = 500$); $\text{CFI}_u^{\text{ref}}$ and $\text{CFI}_u^{\text{asy}}$ = unweighted comparative fit index with reference-state and adapted residual-covariance correction. $n = 301$. The baseline-model SRMR was .285; robust CFI was .677, .858, and .932; Mardia's (1970) multivariate kurtosis was 103.6 against a normal-theory value of 99.

All indices ranked the three models in the same order. The three-factor model fitted best, but its RMSEA was .092, its TLI .896, and residual correlations of $-.19$ (x2–x7) and $-.15$ (x3–x5) remained. Residual correlations between x7 and x8 were .45 and .28 in the simpler models; separating the speed factor reduced this unexplained covariance. These results do not distinguish shared speed ability from a method effect. The corrected indices summarize relative misfit, whereas the individual residuals show which associations remain unexplained.

## Discussion

### *What the Kernel Characterization Adds*

The quadratic-kernel identity locates Lubbe's (2023) unweighted discrepancy within the MMD framework: it compares second moments, not entire distributions. The Gaussian construction has the same large-bandwidth limit but generally a different finite-bandwidth target, and it measures the same residual-correlation misfit through a distance that weights residual directions according to the data distribution. The common-target check shows that the observed nonnormal-data advantage over identically corrected $\mathrm{CFI}_u$ was not solely due to scoring different population quantities. Although Gaussian kernel values are bounded, that property alone does not establish robustness of a procedure that also estimates standardization, correlations, ML parameters, transformations, and bandwidth; the results establish a finite-sample RMSE pattern rather than general robustness.

### *What the Reference State Adds*

The correction extends existing expected-discrepancy bootstrap work (Yuan & Marshall, 2004) by applying a fitted-correlation reference state to paired kernel discrepancies and a comparative ratio. Its value lies in generality: it applies to any residual functional of the fitted correlation structure, including kernel discrepancies for which no analytic sampling correction exists. Where an analytic correction does exist, the two can be compared, and in the primary normal conditions the reference-state and adapted residual-covariance estimators of $\mathrm{CFI}_u$ had RMSEs within .0001. Reference-state correction also improved on ordinary bootstrap correction and the paired U-form estimator in the focused study. These findings support reference-state correction for KCFI within the investigated setting.

The limitation was clearest in heavy-tailed samples at $n = 100$. In correctly specified conditions, the reference-state unweighted correction averaged 98%, 90%, 72%, and 63% of the observed sampling discrepancy under normal, Vale–Maurelli, $t_5$, and contaminated-normal data, respectively, whereas the residual-covariance estimates were 91%–98% (Table S8). The reference-state term also varied little across samples relative to the observed discrepancy, whereas the analytic term tracked it: in the $t_5$ and contaminated-normal cells the correlations between the observed discrepancy and the two terms were .50 and .63 for the reference state and .82 and .80 for the analytic estimator. Under correct specification the ideal correction for a given sample equals its realized sampling discrepancy, so a term that follows the realized value reduces error. The bootstrap term is computed by resampling a sample that already contains the few extreme observations driving the discrepancy, and it cannot reproduce their leverage in the way that a fourth-moment plug-in does; this is consistent with the documented sensitivity of covariance-structure bootstrap estimates to sample size and kurtosis (Yuan & Hayashi, 2006).

***Recommendations and Limitations***

Fit assessment should consider sample-size sensitivity, parsimony, and established interpretive criteria (Hong, 2000; Marsh et al., 1988). Correction reduced sample-size sensitivity here, but the residual-based indices were not uniformly more stable than CFI or robust CFI (Table S3). Neither residual-based index explicitly penalizes model complexity; CFI itself reflects parsimony weakly. Models of different complexity therefore require complementary indices such as TLI and RMSEA or appropriately justified model comparisons (Mulaik et al., 1989). Neither KCFI nor $\mathrm{CFI}_u$ has a well-established cutoff; Lubbe (2023) tentatively suggested .99 for $\mathrm{CFI}_u$. Both depend on the baseline discrepancy, so their values are not absolute misfit quantities comparable across unrelated data sets (van Laar & Braeken, 2022).

The results support three conditional recommendations for continuous-data CFA settings resembling those studied. First, under normal or moderately nonnormal data, the reference-state and analytic corrections of $\mathrm{CFI}_u$ are practically equivalent, and either can supplement conventional fit indices, analysis- and baseline-model SRMRs, and selected large residual correlations. Second, in heavy-tailed small samples, the analytic residual-covariance estimator of $\mathrm{CFI}_u$ should be preferred over both reference-corrected indices. Third, reference-state correction is the tested option for KCFI, whose Gaussian-kernel discrepancy has no analytic correction; when the correction is unavailable because too few bootstrap refits succeed, the analytic $\mathrm{CFI}_u$ estimator and the uncorrected indices should be reported instead. A corrected value of 1 indicates that the estimated discrepancy did not exceed the reference-state correction term, not that the population fits exactly.

The scope is limited to scale-invariant CFA with continuous complete data, one primary model size and loading level, two principal misspecifications, and two severities. Ordered-categorical, missing, multiple-group, and longitudinal data, other residual functions, and robust correlation inputs require separate study. The analytic residual-covariance comparator is an adaptation of the published approach; equivalence under fixed misspecification is not assumed. Results condition on fitting and correction availability, and the empirical example uses a historical sample from two schools that illustrates reporting rather than the indices' behavior in other populations. Vale–Maurelli generation restricts distributional generality (Foldnes & Grønneberg, 2015), the heavy-tailed simulations use only $n = 100$, and bandwidth-selection variability is not separately corrected. Exact kernel evaluation costs $O(n^2p)$ per discrepancy; reference-state correction costs $O(Ln^2p)$ plus model refits and baseline transformations, about 3 s at $n = 200$ with $L = 100$ on one core and 40 s for the empirical model with $L = 500$ (Section S6).

## Conclusion

The unweighted residual discrepancy has an exact quadratic-kernel representation, and Gaussian-kernel KCFI converges to $\mathrm{CFI}_u$ as bandwidth increases. Extending expected-discrepancy bootstrap methods to a fitted-correlation reference state provides a sampling correction for residual functionals that lack an analytic one. In the studied CFA conditions it reduced finite-sample error by 62%–87%, matched the analytic correction of $\mathrm{CFI}_u$ under normality, and improved on ordinary bootstrap correction, while the analytic residual-covariance estimator remained preferable in heavy-tailed small samples. The common-target analysis retained KCFI's nonnormal-data advantage over identically corrected $\mathrm{CFI}_u$. Together, the kernel characterization and the reference-state correction show what residual-based kernel indices measure and when their correction can be trusted.

## Statements and Declarations

### ***Data and Code Availability***

Data, materials, and code are available in Harvard Dataverse (https://doi.org/10.7910/DVN/T6VKHC). The empirical example uses the public Holzinger and Swineford (1939) data distributed with lavaan.

### ***Ethical Considerations***

Ethical approval was not required because this study used simulated data and a publicly available, de-identified secondary data set. The authors did not recruit or interact with participants and had no access to personally identifiable information.

### ***Consent to Participate***

Not applicable.

### ***Consent for Publication***

Not applicable.

### *Competing Interests*

The authors declare no competing interests.

### *Funding*

No funding was received for this study.

### *AI Use Disclosure*

During preparation of this work, the authors used DeepL to translate the initial draft and improve English readability, and OpenAI Codex to assist with refinement and error checking of the research code. The authors reviewed and edited all outputs and take full responsibility for the final manuscript and research code.

## Supporting Information

This supplement provides the computation and proofs, the implemented comparators, sensitivity analyses, and the condition-level comparisons supporting the main text. Complete condition-level outputs are in the analysis package. KCFI = kernel comparative fit index; $\mathrm{CFI}_u$ = unweighted comparative fit index; MMD² = squared maximum mean discrepancy; RMSE = root mean square error; MCSE = Monte Carlo standard error.

## S1. Technical Results

### *S1.1. Population MMD², Bandwidth, and Sample Procedure*

The sample bandwidth is the median of the positive pairwise Euclidean distances in the standardized data. The corresponding population bandwidth $h_P$ is the median of $\|X - X'\|$ for $X, X' \overset{iid}{\sim} P$. In the population analyses, the median from 100,000 independent observation pairs was computed in each of 20 repetitions, and the mean of the repetition-specific medians was used as $\hat{h}_P$; its MCSE was the standard deviation of the 20 medians divided by $\sqrt{20}$. All sample analyses used the exact median of the positive distances among all $n(n-1)/2$ pairs.

For $P = N(0, R_1)$ and $Q = N(0, R_2)$, the population MMD² under the Gaussian kernel $k_h$ has the closed form

$$D_h(R_1, R_2) = \left|I_p + \frac{2R_1}{h^2}\right|^{-1/2} + \left|I_p + \frac{2R_2}{h^2}\right|^{-1/2} - 2\left|I_p + \frac{R_1 + R_2}{h^2}\right|^{-1/2}. \qquad (10)$$

Thus, under normality, $D_M^* = D_{h_P}(R_P, R_M^*)$ and $D_B^* = D_{h_P}(R_P, I_p)$ were computed directly. For nonnormal distributions, $D_M^*$ was estimated by averaging, over independent pairs $X, X' \overset{iid}{\sim} P$, the quantity

$$g_M(X, X') = k_h(X, X') + k_h(A_M^* X, A_M^* X') - k_h(X, A_M^* X') - k_h(A_M^* X, X'), \qquad (11)$$

and $D_B^*$ was computed analogously with $A_B^*$, using the same pairs. Population KCFI MCSEs were computed by a delete-one jackknife over the 20 repetitions.

For each sample, the corrected indices were obtained as follows.

1. Fit the analysis model by ML to the denominator-$n$ covariance matrix, standardize the data to obtain $Z$ and $R$, and rescale the fitted covariance matrix to unit diagonal to obtain $\hat{R}_M$; because the model is scale invariant, fitting to $R$ gives the same $\hat{R}_M$.

2. Compute $A_M$, $A_B = R^{-1/2}$, $Z_M = ZA_M$, $Z_B = ZA_B$, the bandwidth $h$ from $Z$, and the discrepancies $\widehat{D}_M$, $\widehat{D}_B$, $\widehat{U}_M$, and $\widehat{U}_B$.
3. Resample $Z_M$ and $Z_B$ with replacement and restandardize. For the analysis model, refit the model to each bootstrap correlation matrix and transform the bootstrap sample to the refitted structure; for the baseline model, decorrelate each bootstrap sample. The averages of the resulting MMD² and unweighted discrepancies are $\hat{c}_M$, $\hat{c}_B$, $\hat{c}_M^{(\infty)}$, and $\hat{c}_B^{(\infty)}$.
4. Use the corrected values only when the analysis- and baseline-model bootstrap success rates are each at least 90%, and apply the boundary rules of the main text to KCFI and $\mathrm{CFI}_u^{\mathrm{ref}}$.

### *S1.2. Proof of Proposition 1*

Let $P$ and $Q$ be centered distributions with finite second moments and covariance matrices $R_P$ and $R_Q$. For $k_2(x, y) = (x^\top y)^2$ and independent copies $X, X' \sim P$ and $Y, Y' \sim Q$,

$$\mathbb{E}(X^\top X')^2 = \mathbb{E}\{\mathrm{tr}(XX^\top X'X'^\top)\} = \mathrm{tr}(R_P^2), \qquad (12)$$

and similarly $\mathbb{E}(Y^\top Y')^2 = \mathrm{tr}(R_Q^2)$. Independence of $X$ and $Y$ gives $\mathbb{E}(X^\top Y)^2 = \mathrm{tr}(R_P R_Q)$. Substitution into the definition of MMD² yields

$$\mathrm{MMD}^2(P, Q; k_2) = \mathrm{tr}(R_P^2) + \mathrm{tr}(R_Q^2) - 2\mathrm{tr}(R_P R_Q) = \left\|R_P - R_Q\right\|_F^2. \qquad (13)$$

The feature map of $k_2$ contains only the second-order products of the coordinates, so equal embeddings imply equal second moments but not equal distributions; the kernel is therefore not characteristic on general distributions (Sriperumbudur et al., 2011).

### *S1.3. Proof of Theorem 1*

Let $X, X' \overset{iid}{\sim} P$ and $Y, Y' \overset{iid}{\sim} Q$, with the two pairs mutually independent; both distributions have mean zero, finite fourth moments, and covariance matrices $R_P$ and $R_Q$. Write $E = R_P - R_Q$ and $d^2 = \|u - v\|^2$. For each fixed $d$,

$$h^4\left\{\exp\left(-\frac{d^2}{2h^2}\right) - 1 + \frac{d^2}{2h^2}\right\} \to \frac{d^4}{8}. \qquad (14)$$

Because $0 \le e^{-x} - 1 + x \le x^2/2$ for $x \ge 0$, the scaled remainder is bounded by $d^4/8$. Finite fourth moments and the dominated convergence theorem therefore give

$$\mathbb{E}\{k_h(U,V)\} = 1 - \frac{\mathbb{E}\|U-V\|^2}{2h^2} + \frac{\mathbb{E}\|U-V\|^4}{8h^4} + o(h^{-4}), \qquad (15)$$

and the remainder is $O(h^{-6})$ if sixth moments are finite. Substitution into the three expectations defining MMD² cancels the constant terms. Because the means are zero,

$$\mathbb{E}\|X-X'\|^2 = 2\mathrm{tr}(R_P), \qquad \mathbb{E}\|Y-Y'\|^2 = 2\mathrm{tr}(R_Q), \qquad \mathbb{E}\|X-Y\|^2 = \mathrm{tr}(R_P) + \mathrm{tr}(R_Q), \qquad (16)$$

so the $h^{-2}$ terms also cancel. For mutually independent, mean-zero $U$ and $V$ with covariance matrices $\Sigma_U$ and $\Sigma_V$,

$$\mathbb{E}\|U-V\|^4 = \mathbb{E}\|U\|^4 + \mathbb{E}\|V\|^4 + 2\mathrm{tr}(\Sigma_U)\mathrm{tr}(\Sigma_V) + 4\mathrm{tr}(\Sigma_U\Sigma_V), \qquad (17)$$

because the cross terms $\mathbb{E}(\|U\|^2U^\top V)$ and $\mathbb{E}(U^\top V\|V\|^2)$ vanish by independence and zero means. In the combination $\mathbb{E}\|X-X'\|^4 + \mathbb{E}\|Y-Y'\|^4 - 2\mathbb{E}\|X-Y\|^4$, the fourth moments $\mathbb{E}\|X\|^4$ and $\mathbb{E}\|Y\|^4$ cancel exactly, so the fourth-order MMD² term is

$$\frac{1}{8h^4}[2\{\mathrm{tr}(E)\}^2 + 4\|E\|_F^2] = \frac{\{\mathrm{tr}(E)\}^2}{4h^4} + \frac{\|E\|_F^2}{2h^4}. \qquad (18)$$

If $R_P$ and $R_Q$ are correlation matrices, $\mathrm{tr}(E) = 0$, which proves the large-bandwidth limit. With finite sixth moments, $D_M^* = a_M h^{-4} + O(h^{-6})$ and $D_B^* = a_B h^{-4} + O(h^{-6})$; when $a_B > 0$, their ratio is $a_M/a_B + O(h^{-2})$, and $1 - D_M^*/\max(D_M^*, D_B^*) = 1 - a_M/\max(a_M, a_B) + O(h^{-2})$. Thus, KCFI has the stated rate even when the baseline is not the larger discrepancy.

## S2. Population Analyses

### *S2.1. Residual Structures at Equal CFI*

One population contained both a cross-loading and an error covariance, calibrated by bound-constrained nonlinear least squares to .274417 and .341238 so that two analysis models had population CFI = .980: the first estimated the error covariance and fixed the cross-loading to zero, the second did the reverse, and both had the same degrees of freedom.

**Table S1**

*Differences by Residual Structure at Equal Population CFI and ML Misfit*

| **Omitted parameter** | CFI* | $F_M^*$ | $df$ | **SRMR** | **Max** $\lvert r_{\mathrm{res}}\rvert$ | KCFI* | $\mathrm{CFI}_u^*$ |
|---|---|---|---|---|---|---|---|
| Cross-loading | .9800 | .0646 | 86 | .0252 | .0910 | .9853 | .9861 |
| Error covariance | .9800 | .0646 | 86 | .0169 | .0728 | .9931 | .9938 |

*Note.* All values are population values. CFI = comparative fit index; ML = maximum likelihood; $F_M^*$ = population ML discrepancy function value of the analysis model; $df$ = degrees of freedom; SRMR = standardized root mean square residual; Max $|r_{\text{res}}|$ = largest absolute residual correlation; KCFI* = population kernel comparative fit index at the median-heuristic bandwidth; $\text{CFI}_u^*$ = population unweighted comparative fit index. The two analysis models share the same population and baseline model; the common baseline $F_B^*$ is 3.2298.

### *S2.2. Bandwidth and Distributional Sensitivity*

Panel A shows the approach of KCFI to the $\text{CFI}_u$ limit under multivariate normality in the two misspecified populations of the main text (population CFI = .980). Panel B holds $R_P$ and $R_M^*$ fixed across the multivariate normal, Vale–Maurelli (Vale & Maurelli, 1983), multivariate $t_5$, and 5% contaminated-normal distributions (5% of the observations scaled by 3 and the mixture rescaled to unit variance) at common bandwidths estimated under normality.

**Table S2**

*Bandwidth and Distributional Sensitivity of Population KCFI*

*Panel A. Approach to the* $\text{CFI}_u$ *Limit Under Normality*

| **Misspecification** | **Bandwidth** | $\text{KCFI}^*$ | **Difference from limit** |
|---|---|---|---|
| Omitted cross-loading | $0.2h_P$ | .979046 | −.007683 |
| Omitted cross-loading | $0.4h_P$ | .983006 | −.003722 |
| Omitted cross-loading | $h_P$ | .985882 | −.000846 |
| Omitted cross-loading | $2h_P$ | .986505 | −.000223 |
| Omitted cross-loading | $4h_P$ | .986672 | −.000057 |
| Omitted error covariance | $0.2h_P$ | .984132 | −.008945 |
| Omitted error covariance | $0.4h_P$ | .989519 | −.003558 |
| Omitted error covariance | $h_P$ | .992409 | −.000669 |
| Omitted error covariance | $2h_P$ | .992908 | −.000169 |
| Omitted error covariance | $4h_P$ | .993035 | −.000042 |

*Panel B. Population KCFI by Distribution at Common Bandwidths*

| **Misspecification** | **Bandwidth** | **Normal** | **Vale–Maurelli** | $t_5$ | **5% contaminated normal** | **Range** | **Range/MCSE** |
|---|---|---|---|---|---|---|---|
| Omitted cross-loading | $0.5h_P$ | .98399 | .98715 | .98441 | .98465 | .00315 | 93.2 |

| Misspecification | Bandwidth | Normal | Vale–Maurelli | $t_5$ | 5% contaminated normal | Range | Range/MCSE |
|---|---|---|---|---|---|---|---|
| Omitted cross-loading | $h_P$ | .98588 | .98688 | .98595 | .98598 | .00100 | 16.0 |
| Omitted cross-loading | $2h_P$ | .98651 | .98675 | .98666 | .98679 | .00029 | 1.1 |
| Omitted error covariance | $0.5h_P$ | .99063 | .99230 | .99100 | .99123 | .00168 | 39.9 |
| Omitted error covariance | $h_P$ | .99241 | .99280 | .99226 | .99230 | .00054 | 5.9 |
| Omitted error covariance | $2h_P$ | .99291 | .99265 | .99298 | .99244 | .00055 | 1.2 |

*Note.* KCFI* = population kernel comparative fit index; $h_P$ = population median-heuristic bandwidth; MCSE = Monte Carlo standard error. Panel A was computed in closed form (Section S1.1); the $\mathrm{CFI}_u^*$ limits are .986729 for omitted cross-loading and .993078 for omitted error covariance. In Panel B, $h_P$ is the common population bandwidth estimated under normality (5.2133 and 5.2201, respectively); the nonnormal values are estimates from 100,000 independent observation pairs in each of 20 repetitions, and Range/MCSE divides the across-distribution range by the combined MCSE of the maximum and minimum cells.

## S3. Analytic Comparators and Numerical Verification

### *S3.1. Implemented Estimators*

We used Lubbe's (2023) approximation discrepancy ratio (ADR; approach 2), a scaled ADR variant, and an adapted residual-covariance estimator based on approach 1.2 (Maydeu-Olivares, 2017). For $F_{ULS} = \sum_{i<j} e_{ij}^2 = \widehat{U}/2$ and $T_{ULS} = nF_{ULS}$, ADR estimates

$$\hat{\lambda}_u^{\mathrm{ADR}} = \frac{T - df}{T} T_{ULS}. \qquad (19)$$

The baseline term is computed analogously. The scaled variant substitutes $T/\hat{c}$ for $T$, giving the factor $(T - \hat{c}\,df)/T$; we call it scaled ADR without claiming robustness under arbitrary nonnormality. The two corrected discrepancies enter $1 - \max(\hat{\lambda}_{u,M}, 0)/\max(\hat{\lambda}_{u,M}, \hat{\lambda}_{u,B})$ whenever its denominator is positive.

The adapted residual-covariance estimator uses

$$\hat{\lambda}_u^{\mathrm{asy}} = T_{ULS} - \mathrm{tr}(\hat{\Sigma}_e), \qquad \hat{\Sigma}_e = JU\hat{\Gamma}U^\top J^\top, \qquad U = I - \Delta(\Delta^\top W \Delta)^{-1}\Delta^\top W. \qquad (20)$$

Here $s = \mathrm{vech}(S)$; $\hat{\Gamma}$ estimates the asymptotic covariance of $\sqrt{n}(s - \sigma)$ by the distribution-free fourth-moment estimator (Browne, 1984); $\Delta$ is the Jacobian of model-implied covariances; and $J$ maps covariances to correlations, evaluated at $S$. The weight is $W =$ $1/2\, D_p^\top(\hat{\Sigma}^{-1} \otimes \hat{\Sigma}^{-1})D_p$, with $D_p$ the duplication matrix and $\hat{\Sigma}$ the fitted covariance matrix. Diagonal rows of $J$ are zero. For the baseline, $\Delta$ contains derivatives with respect to the $p$ free variances only. The symbol $\Sigma_e$ denotes the sampling covariance of scaled residuals, not a model error-covariance parameter.

This is a covariance-space implementation mapped to correlations, with weights evaluated at the fitted matrix; it is not an exact reproduction of Lubbe's correlation-space implementation, and equivalence under fixed misspecification is not assumed. Evaluating the weight at $S$ instead of $\hat{\Sigma}$ changed the analysis-model correction in $F$ units by .00293–.05085 in one verified replication per focused condition (`table-weight-matrix-sensitivity.csv`). In $F$ units the analytic term is $\mathrm{tr}(\hat{\Sigma}_e)/n$ and the reference term is $\hat{c}_M^{(\infty)}/2$; the factor of two between $\hat{U}$ and $F_{ULS}$ cancels in the comparative ratio. The robust conventional CFI reported in the package uses $T - \hat{c}\, df$ and its baseline analogue (Brosseau-Liard & Savalei, 2014). All covariance and statistic calculations use denominator $n$ and $T = nF_{ML}$.

### *S3.2. Archived Numerical Checks*

Analytic comparators were added by regenerating data from stored seeds. In the primary and severity analyses, the archived ML-statistic check accepted all replications with a stored statistic (9,000 and 5,999). The focused analysis required a converged fit and agreement of both unweighted discrepancies within $|U_{\text{new}} - U_{\text{stored}}| \leq 10^{-6}\max(1, |U_{\text{stored}}|)$; 10,904 of 10,926 analyzed replications passed, and the other 22 (14 of them in the Vale–Maurelli omitted-error-covariance condition at $n = 100$ and CFI = .955) were excluded from the analytic comparisons. Retaining every finite archived comparator value instead would have added 13 reference-eligible replications, changed condition-level RMSE by at most .00029, and changed no KCFI-versus-analytic ordering (`table-exclusion-sensitivity.csv`, `table-regeneration-diagnostics.csv`).

## S4. Finite-Sample and Bootstrap Sensitivity

In the primary simulation, uncorrected bias was negative in every condition (approximately −.071 to −.011), decreasing in magnitude with sample size. Corrected KCFI equaled 1 in 40.8%–51.2% of correctly specified replications; the other corrected indices also showed upper-bound truncation. Full bias, RMSE, and boundary rates are in the simulation summary files.

Table S3, Panel A, gives the range of condition means over $n = 200{,}500{,}1000$, divided by $1 -$ the population index; smaller values indicate less sample-size sensitivity. Panel B reports stronger misspecification at CFI = .955 (cross-loading .39037; error correlation .65497): 5,999 of 6,000 generated samples were successfully analyzed.

**Table S3**

*Sample-Size and Misspecification-Severity Checks*

*Panel A. Relative Range of Means Across Sample Sizes at CFI = .980*

| Distribution / omission | KCFI | $\mathrm{CFI}_u^{\mathrm{ref}}$ | CFI | Robust CFI |
|---|---|---|---|---|
| Normal / Cross | .154 | .183 | .365 | — |
| Normal / Error | .450 | .501 | .427 | — |
| Vale–Maurelli / Cross | .283 | .511 | 1.494 | .663 |
| Vale–Maurelli / Error | .763 | 1.203 | 1.455 | .698 |

*Panel B. RMSE at Population CFI = .955*

| Distribution / omission | $n$ | KCFI | ref | ADR | rADR | asy | Available (%) |
|---|---|---|---|---|---|---|---|
| Normal / Cross | 200 | .01834 | .01777 | .01511 | .01520 | .01777 | 99.2 |
| Normal / Cross | 500 | .00893 | .00855 | .00859 | .00857 | .00854 | 100.0 |
| Normal / Cross | 1,000 | .00636 | .00613 | .00616 | .00614 | .00612 | 100.0 |
| Normal / Error | 200 | .02055 | .02032 | .01659 | .01677 | .02023 | 99.0 |
| Normal / Error | 500 | .01038 | .01043 | .00941 | .00937 | .01041 | 100.0 |
| Normal / Error | 1,000 | .00768 | .00777 | .00715 | .00713 | .00777 | 100.0 |
| Vale–Maurelli / Cross | 200 | .01739 | .01867 | .01819 | .01502 | .01752 | 98.0 |
| Vale–Maurelli / Cross | 500 | .00942 | .00988 | .00897 | .00955 | .00969 | 100.0 |
| Vale–Maurelli / Cross | 1,000 | .00652 | .00682 | .00631 | .00712 | .00681 | 100.0 |
| Vale–Maurelli / Error | 200 | .02210 | .02367 | .02197 | .01816 | .02254 | 95.4 |
| Vale–Maurelli / Error | 500 | .01302 | .01384 | .01199 | .01128 | .01353 | 100.0 |
| Vale–Maurelli / Error | 1,000 | .00815 | .00853 | .00769 | .00820 | .00853 | 100.0 |

*Note.* KCFI = reference-corrected kernel comparative fit index; ref, ADR, rADR, and asy = reference-state, approximation discrepancy ratio, scaled ADR, and adapted residual-covariance estimators of $\mathrm{CFI}_u$; Cross = omitted cross-loading; Error = omitted error covariance. In Panel A,

each entry is the range of the three condition means divided by one minus the population value of the index, so that smaller values indicate greater stability across sample sizes; robust CFI uses the $T - \hat{c}\,df$ correction and is not reported under normality. In Panel B, each RMSE uses its own population target and defined replications; the population $\mathrm{CFI}_u^*$ values are .97403 (cross-loading) and .97007 (error covariance), and Available = reference-state correction available among successfully analyzed samples. Corrected KCFI bias ranged from −.0039 to .0002 and $\mathrm{CFI}_u^{\mathrm{ref}}$ bias from −.0073 to .0003; correction reduced RMSE in all 12 conditions (KCFI, 46.6%–71.7%; $\mathrm{CFI}_u$, 43.0%–68.4%).

In misspecified normal $n = 500$ conditions, replacing $h$ by $0.5h$ or $2h$ changed KCFI by .00145 and .00058 on average (maxima .00560 and .00224). For the bootstrap-setting check, the first 20 samples of each of four misspecified $n = 100$ stress cells were reused (79 of 80 analyzed successfully), and nested prefixes of one $L = 400$ sequence supplied $L = 100$ and 200. At the 90% success criterion, the mean absolute $L = 100$ versus 400 difference was .0030–.0046 for KCFI and .0027–.0038 for $\mathrm{CFI}_u$, or 5.6%–7.0% of between-sample standard deviations (Table S4). Changing the success criterion to 80% or 95% changed only availability (reference 93.7% or 84.8%; ordinary 83.5% or 43.0%), not the value of any estimate that remained eligible.

**Table S4**

*Bootstrap Replications and Availability in the Four Misspecified Stress Cells at a 90% Success Criterion*

| $L$ | Reference available (%) | Ordinary available (%) | Paired samples | Reference RMSE | Ordinary RMSE |
|---|---|---|---|---|---|
| 100 | 88.6 | 69.6 | 55 | .06215 | .07254 |
| 200 | 88.6 | 69.6 | 55 | .06177 | .07295 |
| 400 | 89.9 | 74.7 | 59 | .06930 | .08753 |

*Note.* Reference = reference-state correction; Ordinary = ordinary bootstrap correction; $L$ = number of bootstrap replications. Availability uses the 79 analyzed samples. RMSE uses the samples with both corrections defined at that $L$, so the rows do not use identical subsets. Reference-state correction had lower paired RMSE for both indices at every $L$.

## S5. Focused Comparisons

### *S5.1. Alternatives and Condition-Level Performance*

The paired U-form (diagonal-deleted) estimator is

$$\widehat{D}_U = \frac{1}{n(n-1)} \sum_{i \neq j} \{ k(X_i, X_j) + k(Y_i, Y_j) - k(X_i, Y_j) - k(Y_i, X_j) \}. \quad (21)$$

Unbiasedness holds for fixed kernels and independently sampled pairs, not automatically for the data-dependent standardization and transformation used here. Negative values were retained until the comparative-index boundary rule. Both bootstraps used the original bandwidth and common row indices. Ordinary correction resampled $Z$ and used $2\widehat{D} - \text{mean}_*(\widehat{D}^*)$; reference-state correction resampled $Z_M$ and $Z_B$. Each required at least 90 successful replications out of 100 for both discrepancies.

Among 10,926 successfully analyzed estimated-model samples, reference-state and ordinary corrections were available in 95.2% and 86.3%, respectively; corresponding rates were 99.4%/98.0% in the core, 93.6%/69.2% in the stress, and 84.5%/76.2% in the stronger-misfit groups. Table S5 uses the common-defined subset for each condition.

**Table S5**

*Reference-State Versus Ordinary Bootstrap Correction for KCFI*

| Distribution | Model | $n$ | CFI target | Reference | Ordinary | MSE interval | Available (%) |
|---|---|---|---|---|---|---|---|
| Normal | Exact | 100 | 1.000 | .01868 | .02078 | [.000060, .000105] | 100.0 / 99.0 |
| Normal | Exact | 200 | 1.000 | .01039 | .01045 | [−.000002, .000005] | 100.0 / 100.0 |
| Normal | Cross | 100 | .980 | .02340 | .02533 | [.000062, .000127] | 99.0 / 97.0 |
| Normal | Cross | 200 | .980 | .01404 | .01415 | [−.000003, .000009] | 100.0 / 100.0 |
| Normal | Error | 100 | .980 | .01957 | .02135 | [.000040, .000106] | 99.4 / 98.6 |
| Normal | Error | 200 | .980 | .01198 | .01227 | [.000001, .000013] | 100.0 / 100.0 |
| Vale–Maurelli | Exact | 100 | 1.000 | .02252 | .02768 | [.000179, .000339] | 99.6 / 97.0 |
| Vale–Maurelli | Exact | 200 | 1.000 | .01175 | .01254 | [.000010, .000028] | 100.0 / 100.0 |
| Vale–Maurelli | Cross | 100 | .980 | .02211 | .02559 | [.000120, .000211] | 98.2 / 92.4 |
| Vale–Maurelli | Cross | 200 | .980 | .01461 | .01528 | [.000010, .000030] | 100.0 / 99.8 |
| Vale–Maurelli | Error | 100 | .980 | .02250 | .02676 | [.000159, .000260] | 97.0 / 91.8 |
| Vale–Maurelli | Error | 200 | .980 | .01298 | .01388 | [.000014, .000034] | 100.0 / 100.0 |
| Multivariate $t_5$ | Exact | 100 | 1.000 | .05879 | .07459 | [.000862, .003353] | 94.9 / 74.3 |
| Multivariate $t_5$ | Cross | 100 | .980 | .05093 | .06296 | [.000275, .002465] | 94.7 / 70.9 |
| Multivariate $t_5$ | Error | 100 | .980 | .05073 | .06584 | [.000632, .002891] | 91.5 / 68.6 |
| Contaminated | Exact | 100 | 1.000 | .07265 | .08920 | [.001923, .003437] | 92.9 / 66.1 |
| Contaminated | Cross | 100 | .980 | .06987 | .08433 | [.001570, .002889] | 94.4 / 67.8 |

| Distribution | Model | $n$ | CFI target | Reference | Ordinary | MSE interval | Available (%) |
|---|---|---|---|---|---|---|---|
| Contaminated | Error | 100 | .980 | .07107 | .08524 | [.001581, .002848] | 92.9 / 67.7 |
| Normal | Cross | 100 | .955 | .02443 | .02560 | [.000031, .000085] | 92.1 / 81.9 |
| Normal | Error | 100 | .955 | .02986 | .03360 | [.000135, .000340] | 87.1 / 85.3 |
| Vale–Maurelli | Cross | 100 | .955 | .02523 | .02811 | [.000081, .000226] | 87.0 / 72.7 |
| Vale–Maurelli | Error | 100 | .955 | .02866 | .03385 | [.000210, .000439] | 71.7 / 64.5 |

*Note.* Cross = omitted cross-loading; Error = omitted error covariance; Exact = correct specification. Reference and Ordinary columns are RMSEs against population KCFI. Interval = 95% Monte Carlo interval for ordinary-minus-reference MSE. Available = separate reference/ordinary percentages among analyzed samples; the RMSEs use their intersection. The paired $\mathrm{CFI}_u^{\mathrm{ref}}$ comparisons are in `table-method-performance.csv` and the condition files. On these subsets, KCFI had 10.9%–21.5% lower RMSE than $\mathrm{CFI}_u^{\mathrm{ref}}$ in all 14 nonnormal conditions, relative to each index's own target.

### *S5.2. Analytic Comparisons*

Table S6 uses replications with reference-state correction available, a passed archived verification check, and finite values for all compared estimators; it therefore differs from the ordinary-bootstrap subset in Table S5. Exact counts, biases, and paired Monte Carlo intervals are in `table-focused-comparators.csv`.

**Table S6**

*RMSE of Corrected KCFI and the Unweighted Comparators*

*Panel A. Core Conditions (CFI = .980 When Misspecified)*

| Distribution | Model | $n$ | KCFI | ref | ADR | rADR | asy |
|---|---|---|---|---|---|---|---|
| Normal | Cross | 100 | .02366 | .02315 | .02188 | .02279 | .02335 |
| Normal | Cross | 200 | .01404 | .01336 | .01098 | .01107 | .01325 |
| Normal | Exact | 100 | .01940 | .01982 | .02406 | .02528 | .02032 |
| Normal | Exact | 200 | .01039 | .01002 | .00989 | .01013 | .00997 |
| Normal | Error | 100 | .02007 | .02015 | .02325 | .02434 | .02043 |
| Normal | Error | 200 | .01198 | .01151 | .01112 | .01141 | .01151 |
| Vale–Maurelli | Cross | 100 | .02297 | .02817 | .03655 | .02671 | .02435 |
| Vale–Maurelli | Cross | 200 | .01461 | .01649 | .01814 | .01232 | .01486 |
| Vale–Maurelli | Exact | 100 | .02252 | .02871 | .04164 | .02948 | .02347 |
| Vale–Maurelli | Exact | 200 | .01175 | .01422 | .02053 | .01166 | .01151 |
| Vale–Maurelli | Error | 100 | .02372 | .03016 | .04147 | .03115 | .02536 |

| Distribution | Model | $n$ | KCFI | ref | ADR | rADR | asy |
|---|---|---|---|---|---|---|---|
| Vale–Maurelli | Error | 200 | .01298 | .01472 | .02107 | .01384 | .01271 |

*Panel B. Stress Conditions ($n = 100$; CFI = .980 When Misspecified)*

| Distribution | Model | KCFI | ref | ADR | rADR | asy |
|---|---|---|---|---|---|---|
| Contaminated | Cross | .08966 | .10224 | .12623 | .04284 | .04968 |
| Contaminated | Exact | .09715 | .11110 | .13962 | .04854 | .05223 |
| Contaminated | Error | .08678 | .10315 | .12909 | .04949 | .04860 |
| Multivariate $t_5$ | Cross | .07017 | .08601 | .11186 | .11997 | .05618 |
| Multivariate $t_5$ | Exact | .06805 | .08489 | .11710 | .06880 | .04926 |
| Multivariate $t_5$ | Error | .06859 | .08427 | .11307 | .10852 | .06003 |

*Panel C. Stronger Misspecification ($n = 100$; CFI = .955)*

| Distribution | Model | KCFI | ref | ADR | rADR | asy |
|---|---|---|---|---|---|---|
| Normal | Cross | .02510 | .02441 | .02236 | .02310 | .02461 |
| Normal | Error | .03036 | .03046 | .02658 | .02741 | .03057 |
| Vale–Maurelli | Cross | .02858 | .03156 | .03566 | .02855 | .02938 |
| Vale–Maurelli | Error | .03156 | .03547 | .03786 | .03014 | .03190 |

*Note.* ref, ADR, rADR, and asy are the estimators defined in Section S3.1; asy is the adapted residual-covariance implementation. KCFI targets population KCFI; the other estimators target $\mathrm{CFI}_u^*$. Cross = omitted cross-loading; Error = omitted error covariance; Exact = correct specification. In the six stress conditions, KCFI bias ranged from −.075 to −.036, reference-corrected $\mathrm{CFI}_u$ bias from −.092 to −.055, and asy bias from −.031 to −.023.

### *S5.3. Estimation and Exact-Fit Diagnostics*

**Table S7**

*Interaction Between Model-Correlation Estimation and Correction Method*

| Distribution | Omission | Paired samples | Fixed advantage | Estimated advantage | Interaction | MC interval |
|---|---|---|---|---|---|---|
| Normal | Cross | 485 | −.000205 | .000094 | .000299 | [.000187, .000412] |
| Normal | Error | 493 | −.000051 | .000073 | .000124 | [.000052, .000196] |
| Vale–Maurelli | Cross | 459 | .000309 | .000166 | −.000144 | [−.000303, .000015] |
| Vale–Maurelli | Error | 457 | .000348 | .000210 | −.000138 | [−.000244, −.000032] |

*Note.* The reference advantage is ordinary-minus-reference MSE. The interaction subtracts this advantage in the fixed-correlation condition from that in the estimated-model condition, using the same generated data and common-defined replications. Intervals are 95% Monte Carlo (MC) intervals. Positive values favor reference-state correction or a larger reference advantage under estimation, respectively.

**Table S8**

*Exact-Fit Sampling Discrepancy and Its Two Estimated Corrections in Correctly Specified Conditions*

| Distribution | $n$ | Paired samples | Observed $F_{ULS,M}$: $M$ ($SD$) | Analytic term: $M$ ($SD$) | Analytic (% of observed) | Reference term: $M$ ($SD$) | Reference (% of observed) | $r$(observed, analytic) | $r$(observed, reference) |
|---|---|---|---|---|---|---|---|---|---|
| Normal | 100 | 500 | .606 (.117) | .592 (.035) | 98 | .596 (.036) | 98 | .34 | .29 |
| Vale–Maurelli | 100 | 498 | .705 (.131) | .672 (.056) | 95 | .633 (.042) | 90 | .45 | .30 |
| Multivariate $t_5$ | 100 | 469 | 1.124 (.358) | 1.023 (.248) | 91 | .807 (.084) | 72 | .82 | .50 |
| Contaminated normal | 100 | 460 | 1.261 (.358) | 1.174 (.268) | 93 | .793 (.113) | 63 | .80 | .63 |
| Normal | 200 | 500 | .305 (.057) | .296 (.013) | 97 | .297 (.014) | 97 | .26 | .20 |
| Vale–Maurelli | 200 | 500 | .360 (.066) | .350 (.023) | 97 | .331 (.019) | 92 | .36 | .21 |

*Note.* Under correct specification the population unweighted discrepancy is zero, so the mean observed $F_{ULS,M} = \hat{U}_M/2$ is the exact-fit sampling component on the selected replications. Analytic term = residual-covariance correction $\mathrm{tr}(\hat{\Sigma}_e)/n$; Reference term = reference-state correction $\hat{c}_M^{(\infty)}/2$; $M$ = mean; $SD$ = standard deviation across replications; $r$ = Pearson correlation between the observed discrepancy and the correction term across replications. All quantities use the same reference-eligible, verified replications, and the percentages are ratios of the two means to the observed mean. Because the ideal correction for a sample equals its realized sampling discrepancy, a larger $SD$ and $r$ indicate a term that follows the realized discrepancy more closely.

### *S5.4. Common-Target Sensitivity Analysis*

The primary comparisons score KCFI and $\mathrm{CFI}_u$ against different population quantities. As a sensitivity analysis, we rescored both reference-corrected indices against the same $\mathrm{CFI}_u^*$ value without generating new samples or changing any estimates. Primary and severity comparisons use analyzed samples with reference-state correction available and both estimates finite; focused comparisons additionally require ordinary correction to be available with finite

KCFI, matching Table S5. For each condition, the paired loss contrast is $(\widehat{\mathrm{CFI}}_u^{\mathrm{ref}} - \mathrm{CFI}_u^*)^2 - (\widehat{\mathrm{KCFI}} - \mathrm{CFI}_u^*)^2$; its mean is reported with $\pm 1.96$ times its MCSE, without a multiplicity adjustment.

**Table S9**

*Common-Target Results Across Nonnormal Study Conditions*

| Study | Conditions | Lower RMSE: own targets | Lower RMSE: common target | Positive MC interval | Common-target RMSE reduction (%) |
|---|---|---|---|---|---|
| Primary | 9 | 9 | 9 | 8 | 0.1–16.0 |
| Severity | 6 | 6 | 6 | 6 | 4.2–7.3 |
| Focused | 14 | 14 | 14 | 14 | 11.0–21.2 |
| All | 29 | 29 | 29 | 28 | 0.1–21.2 |

*Note.* Lower RMSE = number of conditions with lower observed KCFI RMSE than reference-corrected $\mathrm{CFI}_u$. Positive MC interval = number of conditions whose entire unadjusted 95% Monte Carlo interval favors KCFI. The 29 study conditions represent 26 distinct sample-size × distribution × misspecification × severity settings because three settings recur in the primary and focused studies. The `diagnostics` command of the analysis package reproduces the table and its condition-level records.

## S6. Empirical and Computing Details

All models were fitted with the analysis package's own Python ML routine; no lavaan (Rosseel, 2012) fit is run by the package, but the Holzinger–Swineford three-factor fit statistics ($\chi^2 = 85.31$, $df = 24$, CFI = .931, TLI = .896, RMSEA = .092, SRMR = .065) reproduce the published lavaan values. The Holzinger–Swineford bandwidth was 3.9007 for all three models. The raw KCFI values were .7226, .8863, and .9380, and raw $\mathrm{CFI}_u$ values were .7486, .9061, and .9478 for the one-, two-, and three-factor models, respectively; the corrected values appear in main Table 3, and complete estimator values and residuals are in the empirical outputs. Marginal skewness ranged from −.3533 to .8666 and excess kurtosis from −.8884 to 1.2398. Mardia's (1970) multivariate kurtosis was 103.59, against the normal-theory value 99 for $p = 9$.

Recorded one-core timings (Python 3.13, one BLAS thread; three repeats) averaged 2.9, 5.8, and 18.1 s for $p = 15$, $L = 100$, and $n = 200{,}500{,}1000$, respectively. The empirical three-factor model at $n = 301$, $p = 9$, and $L = 500$ averaged 39.3 s (36.0–41.4 s). These times include the initial fit, analysis-model bootstrap refits, baseline transformations, and kernel evaluation; full records are in `results/simulation/table-runtime.csv`.

**Supporting References**